\documentclass[twocolumn,showpacs,preprintnumbers,amsmath,amssymb]{revtex4}
\usepackage{amsmath,amssymb,graphics,epsfig,subfigure}
\usepackage{color}

\begin{document}
\newcommand {\nn} {\nonumber}
\renewcommand{\baselinestretch}{1.3}

\title{Relationship between High-Energy Absorption Cross Section and Strong
                   Gravitational Lensing for Black Hole}
\author{Shao-Wen Wei\footnote{E-mail: weishaow06@lzu.edu.cn},
        Yu-Xiao Liu\footnote{E-mail: liuyx@lzu.edu.cn, corresponding author},
        Heng Guo\footnote{E-mail: guoh06@lzu.edu.cn}}
 \affiliation{Institute of Theoretical Physics, Lanzhou University, Lanzhou 730000,
             People's Republic of China}

\begin{abstract}
In this paper, we obtain a relation between the high-energy
absorption cross section and the strong gravitational lensing for a
static and spherically symmetric black hole. It provides us a
possible way to measure the high-energy absorption cross section for a black hole from strong gravitational lensing through astronomical observation. More
importantly, it allows us to compute the total energy emission
rate for high-energy particles emitted from the black hole acting as
a gravitational lens. It could tell us the range of the frequency,
among which the black hole emits the most of its energy and the
gravitational waves are most likely to be observed. We also apply it
to the Janis-Newman-Winicour solution. The results suggest that we can test the
cosmic censorship hypothesis through the observation of
gravitational lensing by the weakly naked singularities acting as
gravitational lenses.
\end{abstract}



\maketitle

A black hole is an object with a strong gravitational field predicted by the general
relativity. Since the spacetime near a black hole is highly curved,
particles---even light---can not escape from the black hole at the
classical level. However, quantum mechanics suggests that black holes
emits radiation like a black body with a finite temperature known as
the Hawking temperature. Therefore, the phenomenon of the absorption
and radiation of gravitational waves in the strong gravitational
field was studied extensively (see, e.g., Refs.
\cite{Matzner,Mashhoon,Starobinsky,Fabbri,Page,Unruh,Sanchez}).

The absorption cross section is one of the essential factors in the
absorption and radiation of gravitational waves. As shown in
\cite{Sanchez}, the total absorption cross section of an ordinary
material sphere monotonically increases with the frequency, while,
for a black hole, the absorption cross section approaches its
constant geometric-optics value in an oscillatory way with the
increasing of the frequency. Thus, by examining the behavior of the
absorption cross section, we can distinguish a black hole from a
material optical absorber. It also can be applied to distinguish
different black holes, since the constant geometric-optics value
varies for different black holes.

A study of the absorption cross section for different black holes of
arbitrary dimension, as well as for all kinds of fields, has been
carried out. At low energy, it was shown that scalars have a cross
section equal to the black hole horizon area and spin-1/2 particles
give the area measured in a flat spatial metric conformally related
to the true metric \cite{DasMaldacena}. At high energy, the absorption cross section oscillates around a limiting constant value. The limiting value was found to be equal to the geometrical cross section of its
photon sphere, which had been explained from the null geodesics and
the analysis of wave theories \cite{Mashhoon,Misner}. In
\cite{Decanini1}, Decanini, Esposito-Farese and Folacci explained
the fluctuations around the limiting value in terms of {black hole
parameters}. Since then, the universal description of the high-energy
absorption cross section for the static and spherically symmetric
black hole has been established.

As stated above, the limiting value of the high-energy absorption
cross section can be interpreted by the null geodesics. On the other
hand, the deflection angle of light in a strong gravitational
lensing is also computed through the null geodesics. As a result, it
allows us to establish a relation between the high-energy
absorption cross section and the strong gravitational lensing (for the gravitational lensing, see Refs.  \cite{Ellis,BozzaReview,Virbhadra,Darwin,Perlickcmp,HassePerlick,Virbhadra2009prd,Chen1}). In fact, there also exists a connection
between the strong gravitational lensing and the quasinormal modes
of spherically symmetric black holes in the eikonal regime, which
was first guessed by Decanini and Folacci \cite{Folacci} and was
realized by Stefanov, Yazadjiev and Gyulchev \cite{Stefanov}, where
they suggested that we could read the gravitational waves we could
expect from a black hole through the
connection established by them. In this current paper, we would like
to go a step further. In the following, we will propose a relation between
the high-energy absorption cross section and the strong
gravitational lensing. From this relation, one can compute the total
energy emission rate for high-energy particles emitted from a black
hole. Then, it could tell us the range of frequency, among which the
gravitational waves are most likely to be observed. We also apply it
to the Schwarzschild black hole solution and the Janis-Newman-Winicour (JNW) solution. The results imply that
we can test the cosmic censorship hypothesis through the observation
of gravitational lensing by the weakly naked singularities, since
there exists thermal radiation for the weakly naked singularities.

Now, let us consider a static and spherically symmetric black hole in
$D(\geq 4)$-dimensional spacetime with the line element assumed as
\begin{eqnarray}
 ds^{2}=-f(r)dt^{2}+\frac{1}{f(r)}dr^{2}+r^{2}d\Omega_{(D-2)}^{2},\label{metric}
\end{eqnarray}
where $d\Omega_{(D-2)}^{2}$ denotes the line element on the unit
$(D-2)$-dimensional sphere $S^{(D-2)}$, for which the usual angular coordinates are $\theta_{i}\in [0,\;\pi]$ $(i=1,...,D-3)$ and $\phi\in [0,\;2\pi]$. The metric function $f(r)$ is imposed with the proper asymptotics  $f({r\rightarrow
+\infty})\rightarrow1$.

Next, we consider a free photon orbiting around a black hole on the
equatorial hyperplane defined by $\theta_{i}=\frac{\pi}{2}$ for
$i=1,...,D-3$. The $t$ motion and $\phi$ motion are, respectively,
associated with the Killing vectors $\partial/\partial t$ and
$\partial/\partial\phi$,
\begin{eqnarray}
 E=f(r)\bigg(\frac{dt}{d\tau}\bigg),\quad
 L=r^{2}\bigg(\frac{d\phi}{d\tau}\bigg),\label{tt}
\end{eqnarray}
where $\tau$ is an affine parameter, and $E$ and $L$ denote the energy
and the orbital angular momentum of the photon, respectively. The
$r$ motion can be expressed as
\begin{eqnarray}
 \bigg(\frac{dr}{d\tau}\bigg)^{2}+V_{\text{eff}}=0,\label{rt}
\end{eqnarray}
with the effective potential $V_{\text{eff}}$ given by
\begin{eqnarray}
 V_{\text{eff}}=\frac{L^{2}}{r^{2}}f(r)-E^{2}. \label{veff}
\end{eqnarray}
Without loss of generality, we set $E=1$. With the null geodesic (\ref{tt})-(\ref{rt}), we could obtain the photon sphere equation for the metric (\ref{metric}). {On the other hand, the existence of the photon sphere in a space-time has important implications for the gravitational lensing (i.e., relativistic images will be produced). So, we would like to give some notes on
the photon sphere. Virbhadra and Ellis, \cite{Ellis} as well as Claudel \emph{et al.} \cite{Claudel}, gave several kinds of definitions of the photon sphere in a static
spherically symmetric spacetime. One definition \cite{Ellis} of the photon sphere is that it is a timelike hypersurface if the Einstein bending angle of a light ray will be unlimited when the closest distance of approach coincides with the photon sphere. An alternative definition was given in \cite{Claudel}, where the photon sphere is well-defined when the spacetime admits a group of symmetries. Especially, Ref. \cite{Claudel} also contains some theorems, which have important implications for astrophysics. These definitions give the same results, and the equivalence can be found in the recent paper \cite{Perlick}. In fact, the photon sphere is known as an unstable circular of light, which provides us a possible way to determine the photon sphere through the effective potential (\ref{veff}).} With detailed analysis, the radius of the photon sphere is found to satisfied the three conditions
\begin{eqnarray}
 V_{\text{eff}}\big|_{r=r_{\text{c}}}=0, \quad
 V_{\text{eff}}'\big|_{r=r_{\text{c}}}=0, \quad
 V_{\text{eff}}''\big|_{r=r_{\text{c}}}<0,
\end{eqnarray}
where $r_{\text{c}}$ is the radius of the photon sphere and the
prime indicates the derivative with respect to $r$. The first condition
admits that the angular momentum
$L=\frac{r_{\text{c}}}{\sqrt{f_{\text{c}}}}$ and the second condition
gives the photon sphere equation for the black hole:
\begin{eqnarray}
 r_{\text{c}}f_{\text{c}}'-2f_{\text{c}}=0.\label{rc}
\end{eqnarray}
The subscript ``c" represents that the metric coefficient $f(r)$ is
evaluated at $r_{\text{c}}$. Solving Eq. (\ref{rc}), we can
determine the radius $r_{\text{c}}$ of the photon sphere. The photon sphere equation is consistent with these obtained by Virbhadra \emph{et al.} through different definitions. The unstable condition of the orbit is shown in the last condition. Combining with the second condition, we have the following unstable condition:
$f_{\text{c}}''-\frac{2}{r_{\text{c}}^{2}}f_{\text{c}}<0$.
It is also clear that, from the last two conditions, the photon sphere is
located at the local maximum of the effective potential.

In the strong deflection limit, Bozza \cite{Bozza} showed that the
deflection angle can be expressed in the form
\begin{eqnarray}
 \alpha(\theta)=-\bar{a}\log\bigg(\frac{\theta D_{\text{OL}}}{u_{\text{c}}}-1\bigg)
                +\bar{b},\label{Atheta}
\end{eqnarray}
where $\theta$ is the angular position of the light source and
$D_{\text{OL}}$ is the observer-lens distance. The minimum impact
parameter $u_{\text{c}}$ and the strong deflection limit
coefficients $\bar{a}$ and $\bar{b}$ are given, in terms of the
metric function $f(r)$, by
\begin{eqnarray}
 u_{\text{c}}&=&L=\frac{r_{\text{c}}}{\sqrt{f_{\text{c}}}}, ~~~~~
 \bar{a}=\sqrt{\frac{2}{2f_{\text{c}}-r_{\text{c}}^{2}f_{\text{c}}''}},\\
 \bar{b}&=&-\pi+b_{\text{R}}+\bar{a}\ln\bigg[\frac{1}{2f_{\text{c}}^{3}}
            \left(\frac{1-f_{\text{c}}}{\bar{a}}\right)^{2}\bigg].
\end{eqnarray}
The expression of the coefficient $b_{\text{R}}$ is in a complicated
form, which can be found in \cite{Bozza}. In order to probe the nature of the lens
from astronomical observations, we should find the relation between
these coefficients and the astronomical observables. For the
purpose of this, we consider the case that the source, lens and observer are
highly aligned. In this case, there is an infinite set of images at
both sides of the black hole. In the simplest situation, we suppose
that the outermost relativistic image with angular position
$\theta_{1}$ is a single image and the rest are packed together
at $\theta_{\infty}$. Then, we have \cite{Bozza}
\begin{eqnarray}
 \theta_{\infty}=\frac{u_{\text{c}}}{D_{\text{OL}}}. \label{LR1}
\end{eqnarray}
The angular separation $s$ and the ratio of the flux $\tilde{r}$ between the first image and the other ones are
\begin{eqnarray}
 s&=&\theta_{1}-\theta_{\infty}=\theta_{\infty}e^{\frac{\bar{b}-2\pi}{\bar{a}}},
   \label{LR2}\\
 \tilde{r}&=&\mu_{1}/\sum_{n=1}^{\infty}\mu_{n}=e^{2\pi/\bar{a}}.\label{LR3}
\end{eqnarray}
The difference between the magnitudes of the outermost relativistic
image and the others  $r_{\text{m}}$ is related to the flux ration
$\tilde{r}$ in the following way: $r_{\text{m}}=2.5\log \tilde{r}$.
Supposing that the value of the distance $D_{\text{OL}}$ is known (later,
we will show that the distance $D_{\text{OL}}$ can also be obtained
through the gravitational lensing), we can obtain the observable
quantities $\theta_{\infty}$, $s$ and $\tilde{r}$ from the
theoretical model. On the other hand, with the data from the
astronomical observation, we can examine the consistency of the
theoretical model and the astronomical observation.

Now, let us turn to the high-energy absorption cross section for
black holes. In \cite{Decanini1}, a universal high-energy absorption
cross section for black holes was presented. Besides the limiting
value, in the eikonal approximation, the fluctuations around the
limiting value are also analyzed using Regge pole techniques.
The compact form of the high-energy absorption cross section for a black
hole with metric (\ref{metric}) was given by \cite{Decanini1}
\begin{eqnarray}
 \sigma_{\text{abs}}(\omega)=\sigma_{\text{lim}}+\sigma_{\text{osc}},\label{ABS}
\end{eqnarray}
with the limiting value $\sigma_{\text{lim}}$ of the absorption cross section $\sigma_{\text{abs}}$ at $\omega\rightarrow \infty$ and the
oscillating part $\sigma_{\text{osc}}$ given by
\begin{eqnarray}
 \sigma_{\text{lim}}&=&\sigma_{\text{geo}}
                 =\frac{\pi^{\frac{D-2}{2}}
                    b_{\text{c}}^{D-2}}{\Gamma(D/2)},\nonumber\\
 \sigma_{\text{osc}}(\omega)&=&
                 (-1)^{D-3}4(D-2)\pi\eta_{\text{c}}e^{-\pi\eta_{\text{c}}}
                 \frac{\sin (\omega T_{\text{c}})}{\omega T_{\text{c}}}\sigma_{\text{geo}}, \nonumber
\end{eqnarray}
where the orbital period $T_{\text{c}}=2\pi b_{\text{c}}$ and
$\sigma_{\text{geo}}$ is the geometrical cross section. And the
other two parameters are
\begin{eqnarray}
 b_{\text{c}}=\frac{r_{\text{c}}}{\sqrt{f_{\text{c}}}},\quad
 \eta_{\text{c}}=\sqrt{f_{\text{c}}-\frac{1}{2}r_{\text{c}}^{2}f_{\text{c}}''}.
\end{eqnarray}
Note that the high-energy absorption cross section
$\sigma_{\text{abs}}$ only depends on the parameters $b_{\text{c}}$
and $\eta_{\text{c}}$. Thus, in order to obtain the relation between the high-energy
absorption cross section and strong gravitational lensing, we just
need to express the parameters of the absorption cross section in terms
of the strong deflection limit coefficients. Comparing these
coefficients together, we get the simple relations:
$b_{\text{c}}=\frac{u_{\text{c}}}{c},~\eta_{\text{c}}=\frac{1}{\bar{a}}$ (the speed of light $c$ has been restored). Furthermore, with the help of (\ref{LR1})-(\ref{LR3}), the absorption cross section coefficients, in
terms of the observables, read
\begin{eqnarray}
 b_{\text{c}}=\frac{D_{\text{OL}}\theta_{\infty}}{c},\quad
 \eta_{\text{c}}=\frac{1}{2\pi}\ln\tilde{r}.\label{bc2}
\end{eqnarray}
Thus, we can rewritten the high-energy absorption cross section
(\ref{ABS}) as
\begin{eqnarray}
 \sigma_{\text{abs}}(\omega)&=&\frac{K\ln \tilde{r}}{\omega \sqrt{\tilde{r}}}\cdot
                   \bigg(\frac{D_{\text{OL}}\theta_{\infty}}{c}\bigg)^{D-3}
                   \sin\Big(\frac{2\pi\omega D_{\text{OL}}\theta_{\infty}}{c}\Big)
                   \nonumber\\
                 &&+\frac{1}{\Gamma(\frac{D}{2})}
                 \bigg(\frac{\sqrt{\pi}D_{\text{OL}}\theta_{\infty}}{c}\bigg)^{D-2},
\end{eqnarray}
with the constant
$K=(-1)^{(D-3)}(D-2)\pi^{(D-4)/2}/\Gamma(\frac{D}{2})$. Here, we
have established the relation between the absorption cross section
and the observables of strong gravitational lensing for black
holes. {This relation admits us to express} the absorption cross section
with these observables of the gravitational lensing.

On the other hand, the gravitational lensing also provides us a
profound way to measure the distance between the black hole lens and
the observer. The method is due to the measurement of time delays
between two consecutive relativistic images. From the method, the
lens-observer distance is \cite{Bozza2}
\begin{eqnarray}
 D_{\text{OL}}=\frac{2\pi c \Delta T_{2,1}}{\theta_{\infty}},
\end{eqnarray}
with $\Delta T_{2,1}$ the time delay between the first relativistic
image and the second one. Substituting it back into (\ref{bc2}), we
obtain $b_{\text{c}}=2\pi\Delta T_{2,1}$. At last, we arrive at the final expression of the high-energy absorption cross section
\begin{eqnarray}
 \sigma_{\text{abs}}(\omega)&=&
         (2\pi\Delta T_{2,1})^{D-2}
         \frac{2\pi K \ln \tilde{r}}{\sqrt{\tilde{r}}}\cdot
                     \frac{\sin (4\pi^{2}\omega\Delta T_{2,1})}
                        {4\pi^{2}\omega\Delta T_{2,1}}\nonumber\\
                     &&+\frac{2^{D-2}\pi^{(3D-6)/2}(\Delta T_{2,1})^{D-2}}{\Gamma(\frac{D}{2})}.
                     \label{lastabs}
\end{eqnarray}

Furthermore, the absorption cross section was also found to be
related to the total energy emitted from the black hole
\cite{Hawking}. For a static and spherically symmetric black hole of
arbitrary dimension, the total energy per unit time and energy
interval $d\omega$ is
\begin{eqnarray}
 \frac{d^{2}E(\omega)}{d\omega dt}
             =\bigg((4\pi)^{\frac{D-1}{2}}\Gamma\big(\frac{D-1}{2}\big)\bigg)^{-1}
              \frac{2\sigma_{\text{abs}}(\omega)}
                {e^{\frac{\omega}{T_{\text{H}}}}-1}\omega^{D-1},\label{energy}
\end{eqnarray}
where $T_{\text{H}}$ is the Hawking temperature of the black hole. Through the detailed analysis, it could tell us the range of the frequency, among which the radiation energy dominates a very large percentage in the total energy.

Here, we would like to study the energy emission rate for the
Schwarzschild black hole solution and the JNW
solution. The JNW solution \cite{Janis} is the most general
static spherically symmetric solution to the Einstein--massless
scalar equations, and its convenient form was presented in \cite{VirbhadraIJMPA}, which reads
\begin{eqnarray}
 ds^{2}&=&\bigg(1-\frac{b}{r}\bigg)^{\nu}dt^{2}-\bigg(1-\frac{b}{r}\bigg)
       ^{-\nu}dr^{2}\nonumber\\
        &&-\bigg(1-\frac{b}{r}\bigg)^{1-\nu}r^{2}(d\vartheta^{2}+\sin^{2}\vartheta d\varphi^{2}), \label{JNW}
\end{eqnarray}
where $ \nu=(1+q^{2})^{-\frac{1}{2}},~b=\sqrt{1+q^{2}}r_{\text{s}}$, $q=\frac{Q}{M}$ is the charge per unit mass $M$ of the black hole,
and the Schwarzschild radius $r_{\text{s}}=2M$. The scalar field of this spacetime reads $\Phi=\frac{qM}{b\sqrt{4\pi}}\ln(1-\frac{b}{r})$.  The solution
(\ref{JNW}) is asymptotically Minkowskian and reduces to the
Schwarzschild solution for $q=0$ or $\nu=1$. As shown in \cite{VirbhadraJoshi},
this solution has a globally naked strong curvature singularity at
$r=b$ for all values of $q\neq 0$ and it satisfies the weak energy
condition. According to \cite{Claudel,VirbhadraKeeton,Virbhadra}, the photon sphere of the JNW spacetime
is
\begin{eqnarray}
 r_{\text{ps}}=\frac{1+2\nu}{2}b
              =(1+\frac{1}{2}\sqrt{1+q^{2}})r_{\text{s}}.
\end{eqnarray}
Since the curvature singularity is at $r=b$, the photon sphere exists
only for $1/2<\nu\leq 1$ (or $0\leq q<\sqrt{3}$). Therefore,
according to the classification presented by Virbhadra \emph{et al.}
\cite{VirbhadraKeeton,VirbhadraNarasimha}, JNW naked singularities are referred to as weakly, marginally, and strongly naked singularities for $0\leq q<\sqrt{3}$, $q=\sqrt{3}$, and
$q>\sqrt{3}$, respectively. In fact, the weakly naked singularities
are those singularities which are contained within at least one
photon sphere, while the marginally and strongly naked ones are those which are not
covered within any photon spheres. Gravitational lensing by the JNW
solution has been studied in \cite{VirbhadraKeeton,VirbhadraNarasimha,Bozza}.
All these results show that the lensing features of weakly
and marginally naked singularities are
qualitatively similar to the Schwarzschild black hole, while the lensing due
to the strongly naked singularities is qualitatively very
different from the Schwarzschild black hole.

\begin{figure}
\subfigure[]{\label{Sigmaabs}
\includegraphics[width=8cm]{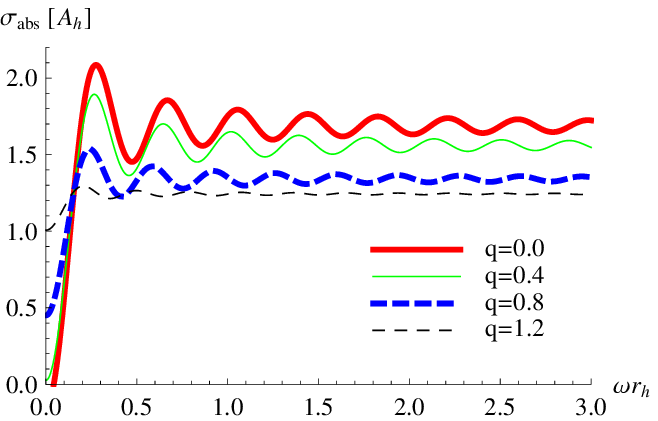}}
\subfigure[]{\label{Sigmalim}
\includegraphics[width=8cm]{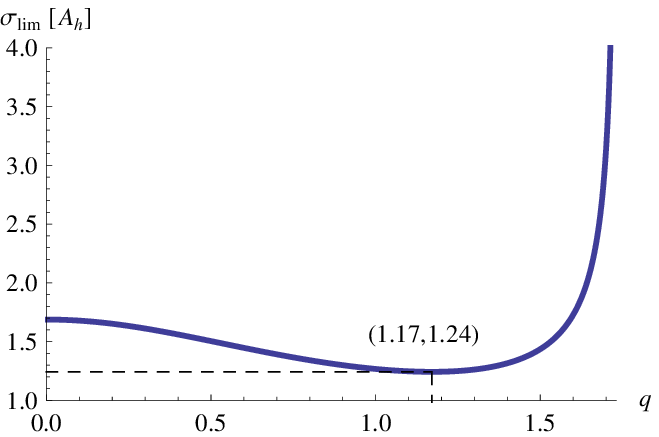}}
\caption{(a) Behavior of the absorption cross section $\sigma_{\text{abs}}$ in units of $A_{\text{h}}=4\pi(1+q^{2})r_{\text{s}}^{2}$. (b) The limiting value $\sigma_{\text{lim}}$ of the absorption cross section in units of $A_\text{h}$ vs charge $q$.}\label{Sigma}
\end{figure}

\begin{figure}
\centerline{
\includegraphics[width=8cm]{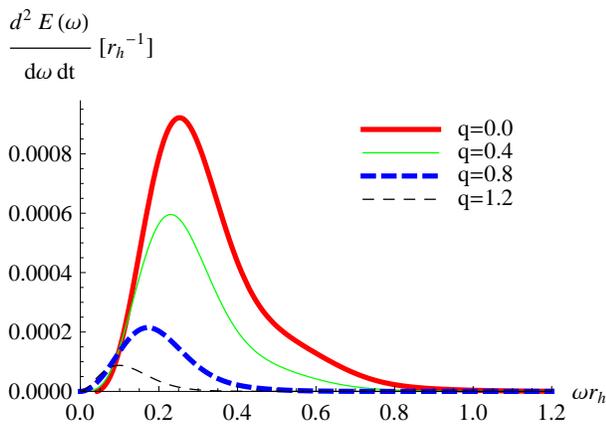}}
\caption{The energy emission rate for the Schwarzschild black hole and the weakly naked singularities.}\label{Erat}
\end{figure}

For the JNW solution, we suppose that the discussion above and
(\ref{ABS}) are held. Then its total high-energy absorption cross
section reads
\begin{eqnarray}
 \sigma_{\text{abs}}=\pi^{2}b_{\text{c}}-\frac{4\pi b_{\text{c}}\eta e^{-\pi \eta} \sin(2\pi b \omega)}{\omega},
\end{eqnarray}
where
\begin{eqnarray}
 b_{\text{c}}&=&{(2+\sqrt{1+q^{2}})}r_{\text{s}}/{2H},\\
 \eta&=&H{\sqrt{q^{4}-4q^{2}+8\sqrt{1+q^{2}}+19}}\Big/{|3-q^{2}|},
\end{eqnarray}
with $H=\big[(q^{2}-4\sqrt{1+q^{2}}+5)/({3-q^{2}})\big]^{\frac{1}{2\sqrt{1+q^{2}}}}$. The total absorption cross section and its limiting values are
described in Fig. \ref{Sigma}. From Fig. \ref{Sigmaabs}, we
can see that the absorption cross section of the JNW solution
oscillates around a limiting constant value, which is just the
geometrical cross section. It was shown in \cite{Decanini1} that, at the high-energy
case, the behavior of the total absorption cross section (\ref{ABS}) is consistent
with the exact one. The limiting values are also found to vary with
the charge density $q$, and the behavior of limiting values is
presented in Fig. \ref{Sigmalim}. We find that the limiting value decreases with the charge density $q$ from 0 to 1.17. When $q$
further increases, $\sigma_{\text{lim}}$ will increase. All these
results describe the Schwarzschild black hole case when $q=0$. It is
also worth noting that these results are for the weakly naked
singularities. For the marginally naked singularities, the
limiting value will be unlimited, and for the strongly naked ones, there is no absorption cross section.
From (\ref{energy}), we could see that the absorption cross section
has an impact on the energy emission rate. Besides it,
the Hawking temperature $T_{\text{H}}$ also has an effect on the energy
emission rate. For the JNW solution, the Hawking temperature
$T_{\text{H}}$ is calculated as
\begin{eqnarray}
 T_{\text{H}}=\frac{1}{4\pi (1+q^{2})r_{\text{s}}}.
\end{eqnarray}
When $q=0$, it describes the temperature of the Schwarzschild black
hole, i.e., $T^{\text{Sch}}_{\text{H}}=\frac{1}{8\pi M}$.
The energy emission rate for it is
\begin{eqnarray}
 \frac{d^{2}E(\omega)}{d\omega dt}
             =\frac{2\pi^{2}\sigma_{\text{abs}}(\omega)}
                {e^{\frac{\omega}{T_{\text{H}}}}-1}\omega^{3}.
\end{eqnarray}
The behavior of the energy emission rate is shown in Fig. \ref{Erat}. The result
displays that, when the charge density $q$ increases, the peak of
$(r_{\text{h}}\frac{d^{2}E(\omega)}{d\omega dt})$ decreases and the
value of $\omega r_{\text{h}}$ corresponding to the peak approaches
zero. These results are for the weakly naked singularities. For the marginally and strongly naked singularities, there is no energy emission rate, which means that there is no thermal radiation for these naked singularities, while there is thermal radiation for the weakly naked singularities. With this result and the relation (\ref{lastabs}), we are allowed to test the cosmic censorship hypothesis through the gravitational lensing by the weakly naked singularities.

At the end of this paper, we would like to make a few comments on
the possible applications of the relation between the parameters of
the strong gravitational lensing and the high-energy absorption
cross section in the eikonal regime presented in the current paper.
It is clear that, with the observation from the strong gravitational
lensing, we can determine the high-energy absorption cross section
and energy emission rate for the gravitational source. The data
analysis of them may provide us with the range of the frequency for
gravitational waves, among which the gravitational source emits the
most of its energy. The gravitational waves are most likely to
be observed in this range. This result may guide us towards detecting the
gravitational waves at a fixed range of the frequency by the
gravitational-wave detectors, such as the Laser Interferometer Space
Antenna and Laser Interferometer Gravitational-wave
Observatory. Furthermore, we also compute the energy emission rate for the Schwarzschild black hole solution and the JNW solution. The results show that, for the weakly naked singularities, there is thermal radiation, which is like that of the Schwarzschild black hole, while, for the marginally and strongly naked singularities, there is no thermal radiation. Thus, these results could provide us with a possible way to test the cosmic censorship hypothesis through the observation of gravitational lensing by the weakly naked singularities acting as gravitational lenses. Another possible application of the result is to determine the dimension $D$ of the spacetime. This may offer us the information about the extra dimensional theories presented in \cite{ADD,RS1}.

The authors are extremely grateful for the anonymous
referee, who suggested that the authors compute and compare results for black holes---weakly, marginally, and strongly naked singularities---and whose suggestion helped the authors obtain an important result.
This work was supported by the Program for New Century Excellent
Talents in University, the Huo Ying-Dong Education Foundation of the
Chinese Ministry of Education (No. 121106), and the National Natural
Science Foundation of China (No. 11075065).

\end{document}